\documentclass[12pt,smallextended,natbib,runningheads,epsfig]{article}
\usepackage{latexsym, amsfonts}
\usepackage{amssymb, amsmath}
\usepackage{amsfonts}
\usepackage{graphicx}
\usepackage{adjustbox}
\usepackage{setspace} 
\usepackage{rotating}
\usepackage{tikz}
\usepackage{bigints}
\usepackage{bm}
\usepackage{multirow}
\usepackage[autostyle]{csquotes}

\newcommand{\be}{\begin{equation}}
\newcommand{\ee}{\end{equation}}
\newcommand{\bea}{\begin{eqnarray}}
\newcommand{\eea}{\end{eqnarray}}
\newcommand{\bean}{\begin{eqnarray*}}
\newcommand{\eean}{\end{eqnarray*}}
\newcommand{\brray}{\begin{array}}
\newcommand{\erray}{\end{array}}
\newcommand{\ben}{\begin{equation}{nonumber}}
\newcommand{\een}{\end{equation}{nonumber}}

\newtheorem{dfn}{Definition}[section]
\newtheorem{thm}[dfn]{Theorem}
\newtheorem{lmma}[dfn]{Lemma}
\newtheorem{ppsn}[dfn]{Proposition}
\newtheorem{crlre}[dfn]{Corollary}
\newtheorem{xmpl}[dfn]{Example}
\newtheorem{rmrk}[dfn]{Remark}

\newcommand{\bdfn}{\begin{dfn}}
\newcommand{\bthm}{\begin{thm}}
\newcommand{\blmma}{\begin{lmma}}
\newcommand{\bppsn}{\begin{ppsn}}
\newcommand{\bcrlre}{\begin{crlre}}
\newcommand{\bxmpl}{\begin{xmpl}}
\newcommand{\brmrk}{\begin{rmrk}}
\newcommand{\edfn}{\end{dfn}}
\newcommand{\ethm}{\end{thm}}
\newcommand{\elmma}{\end{lmma}}
\newcommand{\eppsn}{\end{ppsn}}
\newcommand{\ecrlre}{\end{crlre}}
\newcommand{\exmpl}{\end{xmpl}}
\newcommand{\ermrk}{\end{rmrk}}

\def\a*{{\cal A}_{h,*}}
\def\B{{\cal B}(h)}
\def\B1{{\cal B}_1(h)}
\def\b{{\cal B}^{\rm s.a.}(h)}
\def\b1{{\cal B}^{\rm s.a.}_1(h)}

\numberwithin{equation}{section}

\begin{document}
\begin{center}
{\large {\bf  Testing Multivariate Scatter Parameter in Elliptical Model based on
Forward Search Method}}\\
{\large Chitradipa Chakraborty}\\
{\large IIT Kanpur India }\\
{Email: chitrac@iitk.ac.in}\\
\end{center}

\begin{abstract}
In this article, we establish a test for multivariate scatter parameter in elliptical model, where the location parameter is known, and the scatter parameter is estimated by the multivariate forward search method. The consistency property of the test is also studied here. Inter alia, we investigate the performances of the test for various simulated data, and compare them with those of a classical one.
\vspace{0.1in}

\end{abstract}
%




%

\section{Introduction}

Our objective in this article is to develop a test for multivariate scatter parameter in elliptical distribution, i.e., $H_0 : \Sigma=\Sigma_0$ against $H_1 : \Sigma \neq \Sigma_0$, based on multivariate forward search method. For the purpose, we define first the elliptical distribution of a random variable ${\bf Y}$, the density function of which is $$f_{\bf Y}({\bf y}) = k|\Sigma|^{-\frac{1}{2}} g(({\bf y} -\mbox{\boldmath $\mu$})'\Sigma^{-1}({\bf y} - \mbox{\boldmath $\mu$})).$$
Here, $\mbox{\boldmath $\mu$} \in \mathbb{R}^d$ is the known location parameter, and $\Sigma$ is the $d \times d$ unknown positive-definite scatter matrix. Notation $k$ here is the normalizing constant, i.e., $k = \frac{\Gamma(\frac{d}{2})}{{\pi}^{\frac{d}{2}}}[\int_{0}^{\infty} x^{\frac{d}{2}-1}g(x)dx]^{-1}$, and $g(.)$ is the density generator function such that $\int_{0}^{\infty} x^{\frac{d}{2}-1}g(x)dx < \infty$ (see Fang, Kotz and Ng (1989)). We describe then the multivariate forward search method.\ Though this method is well-known in the literature (see Atkinson, Riani and Cerioli (2009), Johansen and Nielsen (2010)), we here briefly describe the method for the sake of completeness. 

\vspace{0.1in}

The multivariate forward search method is a concept of fitting a model containing outliers to subsets of an increasing size. Given a sample of $n$ observations $\textbf{y}_1,\ldots,\textbf{y}_n$ from an elliptical distribution, the method starts with a subset of cardinality $m$, which is too small in comparison to the original sample size $n$. The unknown parameters of the elliptical distribution are estimated using this subset, and the residuals, or other deviance measures like Mahalanobis distances are computed for all $n$ observations. The subsequent fitting subset is then obtained by taking the $m+h$ observations with the smallest deviance measures for $h \geq 1$. This iteration of fitting and updating scheme continues until all the observations are used in the fitting subset. In practice, $h$ is always a finite number, and its value depends on $n$ and the postulated model. The typical choice, however, is $h = 1$. The estimators of the parameters $\mbox{\boldmath $\mu$}$ and $\Sigma$ at step $\gamma \in (0,1)$ such that $m = [n\gamma]$ are defined as $$\dot{\mbox{\boldmath $\mu$}}_{\gamma,n}=\displaystyle\sum_{i=1}^n \frac{\eta_{i,\gamma,n}}{S_{\gamma,n}}{\bf y}_i ,$$ and $$\dot{\Sigma}_{\gamma,n} = \displaystyle c_\gamma\sum_{i=1}^n \frac{\eta_{i,\gamma,n}}{S_{\gamma,n}}({\bf y}_i-\dot{\mbox{\boldmath $\mu$}}_{\gamma,n})({\bf y}_i-\dot{\mbox{\boldmath $\mu$}}_{\gamma,n})'.$$ Here, $\eta_{i,\gamma,n} = I(Md^2_{i,n} ≤\leq \delta^2_{\gamma,n})$, $S_{\gamma,n} = \displaystyle\sum_{i=1}^n \eta_{i,\gamma,n}$, and $c_\gamma$ is a scaling factor ensuring consistency of $\dot{\Sigma}_{\gamma,n}$, where $I(A) = 1$ if $A$ is true, and otherwise, it equals to zero. Among other notations, $Md^2_{i,n} = ({\bf y}_i-\dot{\mbox{\boldmath $\mu$}}_{\gamma,n})'\Sigma_0^{-1}({\bf y}_i-\dot{\mbox{\boldmath $\mu$}}_{\gamma,n}), \hspace{0.05in}i = 1,...,n$, is the population Mahalanobis distances, where $\Sigma_0$ will be considered as the initial estimator for this forward search methodology at any step, as $\Sigma_0$ is known in null hypothesis.\ Since the multivariate forward search estimator for scatter parameter involves $\dot{\mbox{\boldmath $\mu$}}_{\gamma,n}$, this $\dot{\mbox{\boldmath $\mu$}}_{\gamma,n}$ has been considered as the initial value of the location parameter for computing $Md_{i,n}^2 , i=1,...,n$ (for further details about $\dot{\mbox{\boldmath $\mu$}}_{\gamma,n}$, one may refer to Chakraborty and Dhar (2018+)). Again, $\delta^2_{\gamma,n}$ is the $\gamma$-th quantile among $Md^2_{i,n}, \hspace{0.05in} i = 1,...,n$, and since the observations are obtained from a continuous distribution, $Md^2_{(1),n} < ... < Md^2_{(n),n}$ with probability 1. Thus, we can have $\delta^2_{\gamma,n} = Md^2_{(m),n}$, where $m = [n\gamma]$. Let $H_d(.)$ be the distribution function of the random variable associated with i.i.d squared distances $Md^2_{i,n},\hspace{0.05in} i = 1,....,n$, and the probability density function (denoted as $h_d(.)$) of $Md^2_{i,n}, \hspace{0.05in}i = 1,....,n$ be defined as $$h_d(u )= \displaystyle\frac{\pi^{\frac{d}{2}}}{\Gamma(\frac{d}{2})}u^{\frac{d}{2}-1}g(u) , u\geq 0$$ (see Fang, Kotz and Ng (1989)). The $\gamma$-th quantile of $H_d(.)$ thus can be described as $Q_d(\gamma) = \inf\{x:H_d(x)\geq \gamma\}$, and $c_\gamma = \displaystyle \frac{\gamma}{H_{d+2}(Q_d(\gamma))}$.\ 

\vspace{0.1in}

We know that the sample variance-covariance matrix has asymptotic breakdown point = 0. In this context, we like to point out that the choice of $\gamma = 1/2$ allows the highest possible value of asymptotic breakdown point $= 1/2$ of the multivariate forward search estimator of scatter parameter (see  Section 3). This is one of the significant advantages to constitute the test based on the forward search method. We can now formulate the test statistic $T^1_n=||\sqrt{n}\text{vec}(\dot{\Sigma}_{\gamma,n}-\Sigma_0)||^2$, which is nothing but the square of the Euclidean norm between vec of $\dot{\Sigma}_{\gamma,n}$ and $\Sigma_0$.  Our objective here is to propose a test for the scatter parameter $\Sigma$ using the test statistic $T^1_n$. 

\vspace{0.1in}

In view of the above, our article is organized as follows.\ In Section 2, we have formulated the test statistics based on the multivariate forward search estimator and another classical estimator of the scatter parameter. In that section,
we have also studied the consistency properties of the tests along with their finite sample performances. Section 3 has investigated the robustness property of the estimator. 
Some concluding remarks are given in Section 4. All technical details of the tests are provided in Appendix.

\section{Consistency Property of the Proposed Test and Comparison with the Classical One}

In order to study the consistency property of the test, let us assume that ${\cal{Y}}= \{\textbf{y}_1,\ldots,\textbf{y}_n\}$ is a random sample of size n from an Elliptical distribution,  wherein the scatter parameter $\Sigma$ is unknown, but the location parameter $\mbox{\boldmath $\mu$}$ is known to be  equal to $\dot{\mbox{\boldmath $\mu$}}_{\gamma,n}$. Now, we want to test $H_0 : \Sigma=\Sigma_0$ against the alternative $H_1 : \Sigma \neq \Sigma_0$, where $\Sigma_0$ is specified to us (for further details about testing of hypothesis, see Lehmann and Romano (2005)). To test the above, we formulate a test statistic (denoted as $T^1_n$) based on the multivariate forward search estimator of scatter parameter, as we indicated in the Introduction. We now state a theorem describing the asymptotic behaviour of the test based on $T_n^1$.

\vspace{0.1 in}

\noindent \textbf{Theorem 2.1} {\it Let $c_{\alpha}$ be the $(1 - \alpha)$-th $(0 < \alpha < 1)$ quantile of the distribution of $\sum\limits_{i = 1}^{d^2}\lambda_{i}Z_{i}^{2}$, where, $\lambda_{i}$s are the eigen values of $\displaystyle {c_\gamma^2} (1+\kappa)(I_{d^2}+K_{d,d})(\Sigma_0 \otimes \Sigma_0) + \displaystyle {c_\gamma^2} \kappa \text{vec} \Sigma_0 \text{vec} \Sigma_0'$, and $Z_{i}$s are the i.i.d.\ $N(0, 1)$ random variables. Here, $K_{d,d}$ is a $d^2 \times d^2$ commutation matrix $\displaystyle\sum_{i=1}^d \sum_{i=1}^d J_{ij} \otimes J'_{ij}$, where $J$ is a $d \times d$ matrix with one in the $(i, j)$ position and zeros elsewhere; and $\kappa$ represents the kurtosis of $Y$ defined by $3\kappa = E\left\{(Y_i - \mu_i)^4\right\}/\sigma_{ii} - 3$, where $\mu_i$ is the $i$-th component of $\mbox{\boldmath $\mu$}$ and $\Sigma = ((\sigma_{ij})), i,j = 1,\ldots,n$. A test that rejects $H_0$ when $T^1_n > c_\alpha$, will have asymptotic size $\alpha$. Further, such a test will be a consistent test in the sense that the asymptotic power of the test will be one, when $H_1$ is true.}

\vspace{0.1 in}

\noindent \textbf{Remark 2.1} {\it To implement this test, we have to compute the $(1 - \alpha)$-th quantile of the asymptotic distribution of $||\sqrt{n}\text{vec}(\dot{\Sigma}_{\gamma,n}-\Sigma_0)||^2$, which is nothing but the distribution of $\sum\limits_{i = 1}^{d^2}\lambda_{i}Z_{i}^{2}$. Here, $\lambda_{i}$s are the eigen values of $\displaystyle {c_\gamma^2} (1+\kappa)(I_{d^2}+K_{d,d})(\Sigma_0 \otimes \Sigma_0) + \displaystyle {c_\gamma^2} \kappa \text{vec} \Sigma_0 \text{vec} \Sigma_0'$, and $Z_{i}$'s are the i.i.d.\ $N(0, 1)$ random variables. However, the exact computation of a specified quantile from the aforementioned distribution may not be easily doable. To overcome this issue, one can generate a large sample from the weighted chi-squared distribution and empirically estimate the specified quantile. Also, to compute the power, one should repeatedly generate a large sample from the the weighted chi-squared distribution, where $\lambda_{i}$s are the eigen values of $ \displaystyle {c_\gamma^2} (1+\kappa)(I_{d^2}+K_{d,d})(\Sigma \otimes \Sigma) + \displaystyle {c_\gamma^2} \kappa \text{vec} \Sigma \text{vec} \Sigma'$ for $\Sigma$ $\neq \Sigma_0$. The proportion of $T_{n}^{1} > \hat{c}_{\alpha}$ will be the estimated power, where $\hat{c}_{\alpha}$ is the estimated critical value.} 

\subsection{{\bf Consistency Property of Other Test}}

In this section, we study the consistency properties of the test based on another well-known estimator, as we have already seen that the test based on the forward search estimator is consistent. The sample variance-covariance matrix (denoted as $\hat{S}_n$, most efficient under Gaussian model) is considered to formulate the test statistic. The test statistic for the sample variance-covariance matrix based test is $T^2_n=||\sqrt{n}\text{vec}(\hat{S}_n-\Sigma_0)||^2$, where $\Sigma_0$ is specified in the null hypothesis. In the following proposition, the asymptotic behaviour of the test based on $T^2_n$ is described.

\vspace{0.1 in}

\noindent \textbf{Proposition 2.1} {\it Let $c^\star_{\alpha}$ be the $(1 - \alpha)$-th $(0 < \alpha < 1)$ quantile of the distribution of $\sum\limits_{i = 1}^{d^2}\lambda^\star_{i}Z_{i}^{\star2}$, where, $\lambda^\star_{i}$s are the eigen values of $\displaystyle (1+\kappa)(I_{d^2}+K_{d,d})(\Sigma_0 \otimes \Sigma_0) + \displaystyle  \kappa \text{vec} \Sigma_0 \text{vec} \Sigma_0'$, and $Z^\star_{i}$s are the i.i.d.\ $N(0, 1)$ random variables. Here, $K_{d,d}$ is a $d^2 \times d^2$ commutation matrix $\displaystyle\sum_{i=1}^d \sum_{i=1}^d J_{ij} \otimes J'_{ij}$, where $J$ is a $d \times d$ matrix with one in the $(i, j)$ position and zeros elsewhere; and $\kappa$ represents the kurtosis of $Y$ defined by $3\kappa = E\left\{(Y_i - \mu_i)^4\right\}/\sigma_{ii} - 3$, where $\mu_i$ is the $i$-th component of $\mbox{\boldmath $\mu$}$ and $\Sigma = ((\sigma_{ij})), i,j = 1,\ldots,n$. A test that rejects $H_0$ when $T^2_n > c^{\star}_\alpha$, will have asymptotic size $\alpha$. Further, such a test will be a consistent test in the sense that the asymptotic power of the test will be one, when $H_1$ is true.}

\vspace{0.1 in}


\noindent \textbf{Remark 2.2} {\it Theorem 2.1 and Proposition 2.1 assert that the tests based on $\dot{\Sigma}_{\gamma,n}$ and $\hat{S}_n$ are consistent. In other words, the power of all of them will converge to one as the sample size converges to infinity. Hence, the performances of the tests are comparable when the sample size is infinite.}

\subsection{{\bf Finite sample level and power study}}

In this section, we want to see how the test based on $T^1_n$ performs compared to the test based on $T^2_n$ for the finite sample sizes. 
For this purpose, we consider two distributions, namely, d-dimensional Standard Gaussian distribution and d-dimensional Standard Cauchy distribution with probability density function $f({\bf x}) = \displaystyle \frac{\Gamma\left(\frac{d + 1}{2}\right)}{{\pi}^{\frac{d}{2}}\Gamma\left(\frac{1}{2}\right)}{(1 + {\bf x}^{T}{\bf x})}^{-\frac{d +1}{2}}$. Let us assume $d = 4$ in our numerical study. Under $H_0$, we generate first the data from 4-dimensional standard Gaussian and Cauchy distributions, and to compute the power, we study the distribution of the form $(1-\beta)F + \beta G$, where $\beta \in [0,1]$. 
Here, $F$ is the distribution under $H_0$, i.e., $F({\bf x}) = H({\bf x})$; and $(1-\beta)F + \beta G$ is the distribution under $H_1$, i.e., $G({\bf x}) = |\Sigma|^{-\frac{1}{2}} H(\Sigma^{-\frac{1}{2}}{\bf x})$, where $H$ is any proper distribution function; and $\Sigma$ is the scatter parameter. 
In the first two cases (as shown in Table 1 and Figure 1), we assume $F$ as 4 dimensional standard Gaussian and Cauchy distributions, and  $G$ as 4-dimensional Gaussian and Cauchy distributions with $\Sigma = 5I_4$ and $\gamma = 1/2$. The reason behind this choice of $\gamma$ is that, when $\gamma = 1/2$, the multivariate forward search estimator of scatter parameter attains the highest asymptotic breakdown point (see Section 3). For the last case, i.e., to investigate the robustness property of the test based on $T_{n}^{1}$, we consider $(1-\beta)F + \beta G$ as a mixture of a Gaussian and a point-wise jittered (non-Gaussian, e.g. Cauchy) distribution. Keeping this purpose in mind, we consider 1000 Monte-Carlo replications, each consisting of a sample of size $n = 100$ from alternative distribution with nominal level 5\%. 

\vspace{0.1in}

\noindent \textbf{Table 1:} {Finite sample power of different tests for different values $\beta$ at 5\% level of significance when sample size $ = 100$. Here $\gamma = 1/2$.}
\begin{center}
\begin{adjustbox}{max width=\textwidth} 
\begin{tabular}{ |c|c|c|c|c|c|c|c|c|c|c|c| } 
\hline
Distribution & \multicolumn{11}{|c|}{$H_{0} = N_{4}({\bf 0}, I_{4})$ and $H_{1} =  (1 - \beta)  N_{4}({\bf 0}, I_{4}) + \beta N_{4}({\bf 0}, 5I_{4})$}\\
\hline
$\beta$ & 0 & 0.1 & 0.2 & 0.3 & 0.4 & 0.5 & 0.6 & 0.7 & 0.8 & 0.9 & 1\\
\hline
Test based on $T_{n}^{1}$   & 0.05 & 0.531 & 0.609 & 0.898 & 0.945 & 1 & 1 & 1 & 1 & 1 & 1\\ 
\hline
Test based on $T_{n}^{2}$  & 0.05 & 0.737 & 0.913 & 0.988 & 1 & 1 & 1 & 1 & 1 & 1 & 1\\
\hline
Distribution & \multicolumn{11}{|c|}{$H_{0} = C_{4}({\bf 0}, I_{4})$ and $H_{1} =  (1 - \beta)  C_{4}({\bf 0}, I_{4}) + \beta C_{4}({\bf 0}, 5I_{4})$}\\
\hline
Test based on $T_{n}^{1}$   & 0.05 & 0.634 & 0.78 & 0.883 & 0.999 & 1 & 1 & 1 & 1 & 1 & 1\\ 
\hline
Test based on $T_{n}^{2}$  & 0.05 & 0.196 & 0.213 & 0.202 & 0.18 & 0.556 & 0.02 & 0.123 & 0.288 & 0.304 & 0.327\\
\hline
Distribution & \multicolumn{11}{|c|}{$H_{0} = N_{4}({\bf 0}, I_{4})$ and $H_{1} =  (1 - \beta)  N_{4}({\bf 0}, I_{4}) + \beta C_{4}({\bf 0}, 5I_{4})$}\\
\hline
$\beta$ & 0 & 0.1 & 0.2 & 0.3 & 0.4 & 0.5 & 0.6 & 0.7 & 0.8 & 0.9 & 1\\
\hline
Test based on $T_{n}^{1}$   & 0.05 & 0.593 & 0.667 & 0.885 & 0.919 & 1 & 1 & 1 & 1 & 1 & 1\\ 
\hline
Test based on $T_{n}^{2}$  & 0.2 & 0.797 & 0.811 & 0.925 & 0.936 & 1 & 1 & 1 & 1 & 1 & 1\\
\hline
\end{tabular}
\end{adjustbox}
\end{center}


\vspace{0.2in}



\noindent \textbf{Figure 1:} {Finite sample power of different tests for various values of $\beta$ at 5\% level of significance.}
\begin{center}
\includegraphics[width=5in,height=3.5in]{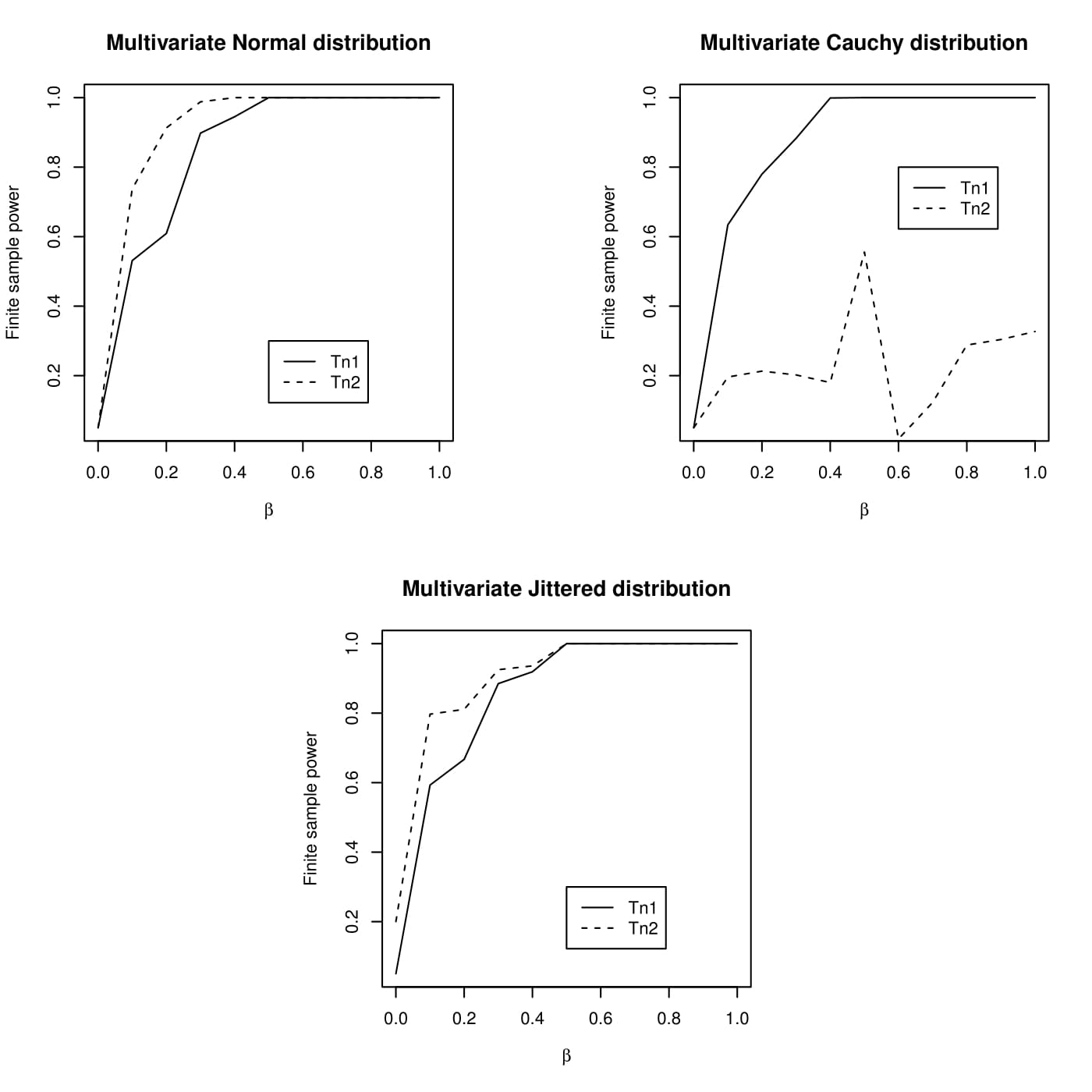}
\end{center}

\vspace{0.1in}
\noindent Then, we compute the powers of the tests based on $T^1_n$ and $T^2_n$ as the proportion of times, the values of the corresponding test statistic exceed the respective critical value. All results are reported in Table 1, and in the table, $N_4$ and $C_4$ denote 4-dimensional Gaussian and Cauchy distributions, respectively.

\vspace{0.1in}

Table 1 and Figure 1 assert that in first two cases, the finite sample power is close to the pre-specified level of the test $ = 0.05$ when $\beta = 0$, since the distribution under $H_1$, i.e., $(1 - \beta) F + \beta G$ coincides with the distribution under the null hypothesis, i.e., $F$. But in the last case, $T_n^2$ exceeds the 5\% level for $\beta = 0$, though $T_n^1$ still is at alpha-level. We have also observed that the test based on $T_{n}^{1}$ performs well compared to the test based on $T_{n}^{2}$ when data are obtained from the heavy-tailed distribution like the mixture of Cauchy distributions.\ This is so since the forward search scatter estimator is more robust than the sample variance-covariance matrix (see Section 3). As it was expected, in the case of Gaussian distribution, the test based on $T^2_n$ (i.e., the test based on the sample variance-covariance matrix) performs better than the tests based on $T^1_n$ (i.e., the test based on the forward search estimator)  since the sample variance-covariance matrix is the maximum likelihood estimator of scatter parameter in the normal distribution. 

\vspace{0.1in}

As we have already observed that the test statistic based on forward search estimator works well when data are generated from the heavy-tailed distribution having outliers, this estimator is expected to be a robust estimator against the presence of outliers in the data. In the view of this, it is our interest now to study the robustness property of the forward search estimator in the next section.
 
\section{Robustness property of ${\dot{\Sigma}}_{\gamma, n}$}

The robustness property of ${\dot{\Sigma}}_{\gamma, n}$ is described by the finite sample breakdown point, which is defined as
$$D\left({\dot{\Sigma}}_{\gamma,n},{\dot{\Sigma}}_{\gamma,n}^{(n^\star)}\right) = \max\left\{\left|\lambda_1({\dot{\Sigma}}_{\gamma,n}) - \lambda_1({\dot{\Sigma}}_{\gamma,n}^{(n^\star)})\right|,\left|\lambda_d({\dot{\Sigma}}_{\gamma,n})^{-1} - \lambda_d({\dot{\Sigma}}_{\gamma,n}^{(n^\star)})^{-1}\right|\right\}.$$ Here, ${\dot{\Sigma}}_{\gamma,n}^{(n^\star)}$ is the forward search estimator of $\Sigma$ computer on a modified sample 
${\cal{Y}}^{(n^\star)}$; and $\lambda_j(.), \hspace{0.05in} j=1,...,d,$ is the $j$-th largest eigenvalue of the corresponding matrix. The finite sample breakdown point of ${\dot{\Sigma}}_{\gamma,n}^{(n^\star)}$ at ${\cal{Y}}$ is defined as $$\epsilon \left({\dot{\Sigma}}_{\gamma,n},{\cal{Y}}\right) = \displaystyle \min_{m^\star\leq n^\star\leq n}\left\{\frac{n^\star}{n}:\sup D\left({\dot{\Sigma}}_{\gamma,n},{\dot{\Sigma}}_{\gamma,n}^{(n^\star)}\right)= \infty\right\},$$ where $m^{*}$ is the cardinality of the initial subset; and ${\dot{\Sigma}}_{\gamma, n}^{(n^\star)}$ is the forward search estimator of $\Sigma$ based on a modified sample ${\cal{Y}}^{(n^\star)} = \{{\bf y}^{*}_{1}, \ldots, {\bf y}^{*}_{n^{*}}, {\bf y}_{n^{*} + 1}, \ldots, {\bf y}_{n}\}$. 
To compute the breakdown point of ${\dot{\Sigma}}_{\gamma, n}$, the following condition is assumed.  

\vspace{0.1 in}

\noindent \textbf{Assumption 1}
{\it Consider $m_0 \geq \left[\frac{n+d+1}{2}\right]$ and denote $\gamma_0 = \displaystyle\frac{m_0}{n}.$  For the initial estimator ${\dot{\Sigma}}_{\gamma_0,n}$, suppose that  $\epsilon \left({\dot{\Sigma}}_{\gamma_0,n}, {\cal{Y}}\right) \geq 1 - \gamma_0.$}

\vspace{0.1 in}

\noindent \textbf{Remark 3.1}
{\it In Assumption 1, the condition on the lower bound of $m_{0}$ ensures that the selected initial observations {\bf y}'s are in the general position, i.e., one cannot draw a hyperplane passing through all the observations, in the case of ${\dot{\Sigma}}_{\gamma_0, n}$. In fact, such condition leads to the highest possible finite sample breakdown point of an affine equivariant estimator (see Davies (1987)). To maintain the highest possible breakdown point, a condition on the lower bound of the breakdown point of  ${\dot{\Sigma}}_{\gamma_0, n}$ is also assumed. We next state the breakdown point of ${\dot{\Sigma}}_{\gamma, n}$, and the treatment here will be parallel to Cerioli, Farcomeni and Riani (2014). }

\vspace{0.1 in}

\noindent \textbf{Theorem 3.1} {\it Suppose that $m > m_0$, $\gamma = \displaystyle\frac{m}{n}$ and Assumption 1 holds. Assume further that {\bf y}'s are in general position as described in Remark 3.1 for $\gamma_{0}$-th step. Then, we have $\epsilon\left({\dot{\Sigma}}_{\gamma, n}, {\cal{Y}}\right) = 1-\gamma.$}

\vspace{0.1 in}

\noindent \textbf{Remark 3.2} {\it Theorem 3.1 asserts that $\gamma$ controls the breakdown point of ${\dot{\Sigma}}_{\gamma, n}$. It is expected that the forward search estimator cannot be {\it breaking down} even in the presence of $(1 - \gamma)$ proportion outliers in the data. The breakdown point of ${\dot{\Sigma}}_{\gamma, n}$ will achieve the highest possible value $1/2$ when $\gamma = 1/2$, and on the other hand, when $\gamma =1\Leftrightarrow m = n$, the breakdown point of the forward search estimator will be equal to $0$.} 

\section{Concluding remarks}
\noindent {\bf Main contribution of the article:} In this article, we have developed a test for multivariate scatter parameter based on the forward search method. We have observed there that our test is consistent when the sample size tends to infinity. For finite sample study also, it possesses good power when data are coming from the heavy-tailed distribution. Overall, our test performs well even in the presence of outliers or large influential observations. This phenomenon is expected because of the robustness property of the multivariate scatter parameter based on the forward search method. However, there is still some scope for further studies under different conditions. We mention below a few of them.    

\vspace{0.1 in}

\noindent {\bf If  $\mbox{\boldmath $\mu$}$ is unknown:} One needs to estimate $\mbox{\boldmath $\mu$}$ when it is unknown, and one may adopt the forward search methodology to estimate $\mbox{\boldmath $\mu$}$.  Given the adaptive nature of the forward search method, when the parameters are unknown, their estimators at step $\gamma$ must be based on the consistent estimators of a previous step. One can take $\gamma_0$, which may represent either the initial step or any step prior to $\gamma$ such that $0<\gamma_0<\gamma<1$. 

\vspace{0.1 in}

\noindent {\bf Non-elliptical distribution:} For non-elliptical distribution, the asymptotic distribution of the forward search estimator for the location parameter remains an open problem. Since the forward search estimator for the scatter parameter depends on the forward search estimator for the location parameter, this extension also remains an open problem, whereas one can extend the results of the test based on sample variance-covariance matrix for non-elliptical distributions under some conditions.


\section{Appendix: Proofs}

\noindent \textbf{Proof of Theorem 2.1:} We have

$\sqrt{n}\text{vec}({\dot{\Sigma}}_{\gamma,n} -\Sigma_0)$\\ 
\noindent $=\sqrt{n}\text{vec}\left(\displaystyle c_\gamma\sum_{i=1}^n \frac{\eta_{i,\gamma,n}}{S_{\gamma,n}}({\bf y}_i-\dot{\mbox{\boldmath $\mu$}}_{\gamma,n})({\bf y}_i-\dot{\mbox{\boldmath $\mu$}}_{\gamma,n})' - \Sigma_0\right)$ \\
\noindent $=\sqrt{n}\text{vec}\left(\displaystyle c_\gamma\sum_{i=1}^n \frac{\eta_{i,\gamma,n}}{S_{\gamma,n}}({\bf y}_i- \mbox{\boldmath $\mu$} + \mbox{\boldmath $\mu$} - \dot{\mbox{\boldmath $\mu$}}_{\gamma,n})({\bf y}_i- \mbox{\boldmath $\mu$} + \mbox{\boldmath $\mu$} - \dot{\mbox{\boldmath $\mu$}}_{\gamma,n})' - \Sigma_0\right)$ \\
\noindent $=\sqrt{n}\text{vec}\left(\displaystyle c_\gamma\sum_{i=1}^n \frac{\eta_{i,\gamma,n}}{S_{\gamma,n}}({\bf y}_i-\mbox{\boldmath $\mu$})({\bf y}_i-\mbox{\boldmath $\mu$})' - \Sigma_0\right)$ (since $\dot{\mbox{\boldmath $\mu$}}_{\gamma,n} \xrightarrow{a.s.} \mbox{\boldmath $\mu$}$)\\
\noindent $=\sqrt{n}\text{vec}\left(\displaystyle \frac{c_\gamma}{S_{\gamma,n}}\sum_{i=1}^n {\eta_{i,\gamma,n}}({\bf y}_i-\mbox{\boldmath $\mu$})({\bf y}_i-\mbox{\boldmath $\mu$})' - \Sigma_0\right)$ \\
\noindent $=\sqrt{n}\displaystyle \frac{c_\gamma}{m} n \text{vec} \left(\displaystyle \frac{1}{n}\sum_{i=1}^n {\eta_{i,\gamma,n}}({\bf y}_i-\mbox{\boldmath $\mu$})({\bf y}_i-\mbox{\boldmath $\mu$})' - \displaystyle \frac{m}{nc_\gamma}\Sigma_0\right)$ (since $ S_{\gamma,n} = m$)\\
\noindent $=\sqrt{n}\displaystyle {c_\gamma}\gamma^{-1}\text{vec} \left(\displaystyle \frac{1}{n}\sum_{i=1}^n {\eta_{i,\gamma,n}}({\bf y}_i-\mbox{\boldmath $\mu$})({\bf y}_i-\mbox{\boldmath $\mu$})' - \displaystyle \gamma c_\gamma^{-1}\Sigma_0\right)$ (since $\displaystyle \frac{m}{n} = \gamma$)

\vspace{0.1in}

\noindent As we have $E\left({\eta_{i,\gamma,n}}({\bf y}_i-\mbox{\boldmath $\mu$})({\bf y}_i-\mbox{\boldmath $\mu$})'\right) = \gamma c_\gamma^{-1}\Sigma_0$, to derive the distribution of $\left(\sqrt{n}\text{vec}({\dot{\Sigma}}_{\gamma,n} -\Sigma_0)\right)$, we need to compute $Var(\text{vec}({\eta_{i,\gamma,n}}({\bf y}_i-\mbox{\boldmath $\mu$})({\bf y}_i-\mbox{\boldmath $\mu$})'))$. Suppose that $\text{vec}({\bf y}_i-\mbox{\boldmath $\mu$})({\bf y}_i-\mbox{\boldmath $\mu$})' = {\bf X}_i$, where ${\bf X}_i= (X_{11,i},X_{21,i},\ldots,X_{d1,i},\\ X_{12,i},X_{22,i},\ldots,X_{d2,i},\ldots,X_{1d,i},X_{2d,i},\ldots,X_{dd,i})'$ is a $d^2 \times 1$ vector. We now have to find $cov(\eta_jX_{1j,i}, \eta_kX_{1k,i})$. Observe that,

$cov(\eta_jX_{1j,i}, \eta_kX_{1k,i})$\\
\noindent $=E[\eta_j(X_{1j,i}-E(X_{1j,i}))\eta_k(X_{1k,i}-E(X_{1j,i}))]$\\
\noindent $=P(\eta_j=1,\eta_k=1)E[(X_{1j,i}-E(X_{1j,i}))(X_{1k,i}-E(X_{1j,i}))]$\\
\noindent $=\gamma^2E[(X_{1j,i}-E(X_{1j,i}))(X_{1k,i}-E(X_{1j,i}))]$

\vspace{0.1in}

\noindent This implies $Var(\text{vec}({\eta_{i,\gamma,n}}({\bf y}_i-\mbox{\boldmath $\mu$})({\bf y}_i-\mbox{\boldmath $\mu$})'))= \gamma^2 V(\text{vec}(({\bf y}_i-\mbox{\boldmath $\mu$})({\bf y}_i-\mbox{\boldmath $\mu$})'))$

\vspace{0.1in}

\noindent Therefore, 

\noindent $Var(\text{vec}({\eta_{i,\gamma,n}}({\bf y}_i-\mbox{\boldmath $\mu$})({\bf y}_i-\mbox{\boldmath $\mu$})'))= \gamma^2\left[(1+\kappa)(I_{d^2}+K_{d,d})(\Sigma_0 \otimes \Sigma_0) + \kappa \text{vec} \Sigma_0 \text{vec} \Sigma_0'\right]$ (see Tyler (1982))

\vspace{0.1in}

\noindent This implies that

\noindent $Var(\text{vec}(\sqrt{n}({\dot{\Sigma}}_{\gamma,n} -\Sigma_0))= {c_\gamma}^2\left[(1+\kappa)(I_{d^2}+K_{d,d})(\Sigma_0 \otimes \Sigma_0) + \kappa \text{vec} \Sigma_0 \text{vec} \Sigma_0'\right]$. 

\vspace{0.1in}

To test $H_{0}: \Sigma = \Sigma_0$ against $H_{1}: \Sigma\neq \Sigma_0$, the power of the test based on $T_{n}^{1}$ is given by $P_{H_{1}}\left[T_{n}^{1} > c_{\alpha}\right]$, where $c_{\alpha}$ is the $(1 - \alpha)$-th $(0 < \alpha < 1)$ quantile of the distribution of $\sum\limits_{i = 1}^{d^2}\lambda_{i}Z_{i}^{2}$. Here, $\lambda_{i}$s are the eigen values of $\displaystyle {c_\gamma^2} (1+\kappa)(I_{d^2}+K_{d,d})(\Sigma_0 \otimes \Sigma_0) + \displaystyle {c_\gamma^2} \kappa \text{vec} \Sigma_0 \text{vec} \Sigma_0'$, and $Z_{i}$s are the i.i.d.\ $N(0, 1)$ random variables. In view of the orthogonal decomposition of multivariate normal distribution, $T_{n}^{1}$ converges weakly to the distribution of  $\sum\limits_{i = 1}^{d^2}\lambda_{i}Z_{i}^{2}$, and hence, the asymptotic size of the test based on $T_{n}^{1}$ is $\alpha$. Let us now denote $\Sigma = \Sigma_1 (\neq \Sigma_0)$ under $H_{1}$, and we now consider

\vspace{0.2 in}

$\displaystyle\lim_{n\rightarrow\infty} P_{H_1} \left[T_{n}^{1} > c_{\alpha}\right]$ \\$= \displaystyle\lim_{n\rightarrow\infty} P_{H_1} \left[\left|\left|\sqrt{n}\text{vec}({\dot{\Sigma}}_{\gamma,n} -\Sigma_0)\right|\right|^2 > c_{\alpha}\right]$\\ $= \displaystyle\lim_{n\rightarrow\infty} P_{H_1} \left[\left|\left|\sqrt{n}\text{vec}({\dot{\Sigma}}_{\gamma,n} - \Sigma_1 + \Sigma_1 - \Sigma_0)\right|\right|^2 > c_{\alpha}\right]$\\$ = \displaystyle\lim_{n\rightarrow\infty} P_{H_1} \left[\left|\left|\sqrt{n} \text{vec} ({\dot{\Sigma}}_{\gamma,n} - \Sigma_{1})\right|\right|^2 + \left|\left|\sqrt{n}\text{vec}(\Sigma_{1}-\Sigma_{0})\right|\right|^2 + 2 \left<\sqrt{n} \text{vec}({\dot{\Sigma}}_{\gamma,n} - \Sigma_{1}), \sqrt{n}\text{vec}(\Sigma_{1}-\Sigma_{0}) \right>  > c_{\alpha}\right]$ \\$=\displaystyle\lim_{n\rightarrow\infty} P_{H_1} \left[\left|\left|\sqrt{n} \text{vec}({\dot{\Sigma}}_{\gamma,n} - \Sigma_1)\right|\right|^2 > c_{\alpha} - \left|\left|\sqrt{n}\text{vec}(\Sigma_{1}-\Sigma_{0})\right|\right|^2 - \\2n \left<\text{vec}({\dot{\Sigma}}_{\gamma,n} - \Sigma_{1}), \text{vec}(\Sigma_{1}-\Sigma_{0}) \right>\right]$ \\$= \displaystyle\lim_{n\rightarrow\infty}P_{H_1} \left[\left|\left|\sqrt{n} \text{vec}({\dot{\Sigma}}_{\gamma,n} - \Sigma_1)\right|\right|^2  > c_{\alpha} - n\left|\left|\text{vec}(\Sigma_{1}-\Sigma_{0})\right|\right|^2 \right]$ 
since under $H_{1}$, ${\dot{\Sigma}}_{\gamma,n} \xrightarrow{a.s.} \Sigma_{1}$ \\$ \rightarrow 1 $ as $n\rightarrow\infty$. 

\vspace{0.1in}

The last implication follows from the fact that $\left|\left|\sqrt{n} \text{vec}({\dot{\Sigma}}_{\gamma,n} - \Sigma_1)\right|\right|^2$ converges weakly to  the distribution of $\sum\limits_{i = 1}^{d^2}\lambda_{i}Z_{i}^{2}$ under $H_{1}$, and $c_{\alpha} - n\left|\left|\text{vec}(\Sigma_{1}-\Sigma_{0})\right|\right|^2 $ converges to $-\infty$ as $n\rightarrow\infty$. This fact leads to the result. Hence the proof is complete. $\hfill\Box$

\noindent \textbf{Proof of Proposition 2.1:} We have

$\sqrt{n}\text{vec}(\hat{S}_n-\Sigma_0)$\\
\noindent $=\sqrt{n}\text{vec}\left(\displaystyle \frac{1}{n}\sum_{i=1}^n ({\bf y}_i- \bar{{\bf y}})({\bf y}_i-\bar{{\bf y}})' - \Sigma_0\right)$ \\
\noindent $=\sqrt{n}\text{vec}\left(\displaystyle \frac{1}{n}\sum_{i=1}^n ({\bf y}_i- \mbox{\boldmath $\mu$} + \mbox{\boldmath $\mu$} - \bar{{\bf y}})({\bf y}_i- \mbox{\boldmath $\mu$} + \mbox{\boldmath $\mu$} - \bar{{\bf y}})' - \Sigma_0\right)$ \\
\noindent $=\sqrt{n}\text{vec}\left(\displaystyle \frac{1}{n}\sum_{i=1}^n ({\bf y}_i-\mbox{\boldmath $\mu$})({\bf y}_i-\mbox{\boldmath $\mu$})' - \Sigma_0\right)$ (since $\bar{{\bf y}} \xrightarrow{a.s.} \mbox{\boldmath $\mu$}$)

\vspace{0.1in}

\noindent As we have $E\left(({\bf y}_i-\mbox{\boldmath $\mu$})({\bf y}_i-\mbox{\boldmath $\mu$})'\right) = \Sigma_0$, to derive the distribution of $\text{vec}\left(\sqrt{n}(\hat{S}_n-\Sigma_0)\right)$, we need to compute $Var(\text{vec}(({\bf y}_i-\mbox{\boldmath $\mu$})({\bf y}_i-\mbox{\boldmath $\mu$})'))$. From Tyler (1982), we can have 

\vspace{0.1in}

\noindent $Var(\text{vec}(({\bf y}_i-\mbox{\boldmath $\mu$})({\bf y}_i-\mbox{\boldmath $\mu$})'))= (1+\kappa)(I_{d^2}+K_{d,d})(\Sigma_0 \otimes \Sigma_0) + \kappa \text{vec} \Sigma_0 \text{vec} \Sigma_0'$. 

\vspace{0.1in}

To test $H_{0}: \Sigma = \Sigma_0$ against $H_{1}: \Sigma\neq \Sigma_0$, the power of the test based on $T_{n}^{2}$ is given by $P_{H_{1}}\left[T_{n}^{2} > c^\star_{\alpha}\right]$, where $c_{\alpha}$ is the $(1 - \alpha)$-th $(0 < \alpha < 1)$ quantile of the distribution of $\sum\limits^{d^2}_{i = 1} \lambda^\star_{i}Z_{i}^{\star2}$. Here, $\lambda^\star_{i}$s are the eigen values of $\displaystyle (1+\kappa)(I_{d^2}+K_{d,d})(\Sigma_0 \otimes \Sigma_0) + \displaystyle \kappa \text{vec} \Sigma_0 \text{vec} \Sigma_0'$, and $Z^\star_{i}$s are the i.i.d.\ $N(0, 1)$ random variables. In view of the orthogonal decomposition of multivariate normal distribution, $T_{n}^{2}$ converges weakly to the distribution of  $\sum\limits_{i = 1}^{d^2}\lambda^\star_{i}Z_{i}^{\star2}$, and hence, the asymptotic size of the test based on $T_{n}^{2}$ is $\alpha$. Let us now denote $\Sigma = \Sigma_1 (\neq \Sigma_0)$ under $H_{1}$, and we now consider


$\displaystyle\lim_{n\rightarrow\infty} P_{H_1} \left[T_{n}^{2} > c^\star_{\alpha}\right]$ \\$= \displaystyle\lim_{n\rightarrow\infty} P_{H_1} \left[\left|\left|\sqrt{n}\text{vec}(\hat{S}_n -\Sigma_0)\right|\right|^2 > c^\star_{\alpha}\right]$\\ $= \displaystyle\lim_{n\rightarrow\infty} P_{H_1} \left[\left|\left|\sqrt{n}\text{vec}(\hat{S}_n - \Sigma_1 + \Sigma_1 - \Sigma_0)\right|\right|^2 > c^\star_{\alpha}\right]$\\$ = \displaystyle\lim_{n\rightarrow\infty} P_{H_1} \left[\left|\left|\sqrt{n} \text{vec} (\hat{S}_n - \Sigma_{1})\right|\right|^2 + \left|\left|\sqrt{n}\text{vec}(\Sigma_{1}-\Sigma_{0})\right|\right|^2 + 2 \left<\sqrt{n} \text{vec}(\hat{S}_n - \Sigma_{1}), \sqrt{n}\text{vec}(\Sigma_{1}-\Sigma_{0}) \right>  > c^\star_{\alpha}\right]$ \\$=\displaystyle\lim_{n\rightarrow\infty} P_{H_1} \left[\left|\left|\sqrt{n} \text{vec}(\hat{S}_n - \Sigma_1)\right|\right|^2 > c^\star_{\alpha} - \left|\left|\sqrt{n}\text{vec}(\Sigma_{1}-\Sigma_{0})\right|\right|^2 - 2n \left<\text{vec}(\hat{S}_n - \Sigma_{1}), \text{vec}(\Sigma_{1}-\Sigma_{0}) \right>\right]$ \\$= \displaystyle\lim_{n\rightarrow\infty}P_{H_1} \left[\left|\left|\sqrt{n} \text{vec}(\hat{S}_n - \Sigma_1)\right|\right|^2  > c^\star_{\alpha} - n\left|\left|\text{vec}(\Sigma_{1}-\Sigma_{0})\right|\right|^2 \right]$ 
since under $H_{1}$, $\hat{S}_n \xrightarrow{a.s.} \Sigma_{1}$ \\$ \rightarrow 1 $ as $n\rightarrow\infty$. 

\vspace{0.1in}

The last implication follows from the fact that $\left|\left|\sqrt{n} \text{vec}(\hat{S}_n - \Sigma_1)\right|\right|^2$ converges weakly to  the distribution of $\sum\limits_{i = 1}^{d^2}\lambda^\star_{i}Z_{i}^{\star2}$ under $H_{1}$, and $c^\star_{\alpha} - n\left|\left|\text{vec}(\Sigma_{1}-\Sigma_{0})\right|\right|^2 $ converges to $-\infty$ as $n\rightarrow\infty$. This fact leads to the result. Hence the proof is complete. $\hfill\Box$

\vspace{0.1in}

\noindent \textbf{Proof of Theorem 3.1:} Suppose that the original observations are denoted by ${\cal{Y}} = \{{\bf y}_{1}, \ldots, {\bf y}_{n}\}$, and without loss of generality, first $n^{*} < n$ observations are corrupted. Let ${\cal{Y}}^{*} = \{{\bf y}_{1}^{*}, \ldots, {\bf y}_{n^{*}}^{*}, {\bf y}_{n^{*} + 1}, \ldots, {\bf y}_{n}\}$ denote the contaminated sample, where ${\bf y}_{i}^{*}$ are corrupted observations, $i = 1, \ldots, n^{*}$. For any $d$-dimensional vector $b \neq 0$, $$\lambda_1({\dot{\Sigma}}_{\gamma,n})= \sup_b \frac{b'{\dot{\Sigma}}_{\gamma,n}^{-1}b}{b'b}.$$ It follows that $$\sup\left\{\left|\lambda_1({\dot{\Sigma}}_{\gamma,n}) - \lambda_1({\dot{\Sigma}}_{\gamma,n}^{(n^\star)})\right|\right\}=\infty$$ if and only if $\left|\left|{\bf y}_{i}^\star\right|\right|=\infty$ for any $i = 1, \ldots, n^{*}$. Similarly, under Assumption 1, $$\sup\left\{\left|\lambda_v({\dot{\Sigma}}_{\gamma,\gamma_0,n})^{-1} - \lambda_v({\dot{\Sigma}}_{\gamma,\gamma_0,n}^{(n^\star)})^{-1}\right|\right\}=\infty$$ for atleast $n^\star=m-d$ units. This fact implies that either ${\bf y}_{i}^\star \propto (1,....,1)'$, or $\exists l \in \{n^\star +1,...,n\}$ such that ${\bf y}_{i}^\star \propto {\bf y}_l$. Without loss of generality, suppose that the aforementioned equivalent relationship holds for some $k$, and we than have $\eta_{k,\gamma,n} = 1$ since $\sum\limits_{i = 1}^{n}\eta_{i,\gamma, n} = m>0$. This fact implies that $Md^2_{k,n} = \infty$ for those choices of $k$. Let us further consider that there are $n_{1}^{*} < n^{*}$ many choices of $k$ for which $Md^2_{k,n} = \infty$. Now, in view of the definition of $\eta_{k,\gamma, n}$, we have $\eta_{j,\gamma,n} = I(Md^2_{j,n} \leq \delta^2_{\gamma, n}) = I(Md^2_{j, n} \leq Md^2_{(m), n}) = 0$ for $j = 1, \ldots, n_{1}^{*}$ when $n_{1}^{*} < n - m$. Further, note that $Md^2_{k,n} = \infty$ for any $k = 1, \ldots, n^{*}$ when $n_{1}^{*} = n^{*}$. Hence the proof is complete.

\vspace{0.1in}

\noindent {\bf Acknowledgment:} The author is thankful to the Co-Editor-in-Chief\ Prof. Yimin Xiao, an anonymous Associate Editor and an anonymous reviewer for their valuable suggestions that have improved the article significantly. The author would also like to thank her Ph.D. supervisor Dr. Subhra Sankar Dhar for his critical comments and suggestions to refine the article remarkably.




\end{document}